\documentclass[aps,prb,twocolumn,amsmath,amssymb,superscriptaddress,reprint,longbibliography]{revtex4-2}

\usepackage[utf8]{inputenc}
\usepackage{graphicx}
\usepackage[colorlinks=true,citecolor=blue]{hyperref}
\usepackage{units}

\newcommand{\bra}[1]{\langle #1|}
\newcommand{\ket}[1]{|#1\rangle}

\DeclareMathOperator{\sign}{sign}
\newcommand{\up}{\uparrow}
\newcommand{\down}{\downarrow}
\renewcommand{\vec}[1]{\mathbf{#1}}

\newcommand{\kBT}{k_\text{B}T}
\DeclareMathOperator{\re}{Re}
\DeclareMathOperator{\im}{Im}

\begin{document}
\title{Higgs-like pair amplitude dynamics in superconductor-quantum dot hybrids}
\author{Mathias Kamp}
\affiliation{Theoretische Physik, Universität Duisburg-Essen and CENIDE, D-47048 Duisburg, Germany}
\author{Björn Sothmann}
\affiliation{Theoretische Physik, Universität Duisburg-Essen and CENIDE, D-47048 Duisburg, Germany}
\date{\today}

\begin{abstract}
We consider a quantum dot weakly tunnel coupled to superconducting reservoirs. A finite superconducting pair amplitude can be induced on the dot via the proximity effect. We investigate the dynamics of the induced pair amplitude after a quench and under periodic driving of the system by means of a real-time diagrammatic approach. We find that the quench dynamics is dominated by an exponential decay towards equilibrium.
In constrast, the periodically driven system can sustain coherent oscillations of both the amplitude and the phase of the induced pair amplitude in analogy to Higgs and Nambu-Goldstone modes in driven bulk superconductors.
\end{abstract}

\maketitle
\section{\label{sec:intro}Introduction}
Superconductivity has been an active field of research since its discovery more than one hundred years ago. From a fundamental point of view, it constitutes a macroscopic manifestation of quantum coherence that gives rise to interesting phenomena such as flux quantization in superconducting rings~\cite{doll_experimental_1961,deaver_experimental_1961,byers_theoretical_1961} and the Josephson effect~\cite{josephson_possible_1962}, i.e., the dissipationless flow of charge currents in superconducting junctions in the absence of any bias voltage.
At the same time it is also of relevance for applications such as superconducting quantum interference devices~\cite{jaklevic_quantum_1964} that can act as extremely sensitive magnetometers.

A microscopic understanding of superconductivity has been achieved within BCS theory~\cite{bardeen_theory_1957} which describes the transition between a normal metal and a superconductor as a second-order phase transition in which electrons condense into $s$-wave, spin-singlet Cooper pairs. The associated superconducting order parameter is given by the macroscopic wave function of the Cooper pairs $\Delta e^{i\phi}$ whose form indicates the breaking of the U(1) symmetry. This breaking of a continuous symmetry implies the existence of collective gapless excitations. For a superconductor, these Nambu-Goldstone modes correspond to fluctuations of the phase of the superconducting order parameter. While a superconductor is invariant under a variation of the phase, the Nambu-Goldstone modes are shifted to the plasma frequency by the Anderson-Higgs mechanism~\cite{anderson_coherent_1958,nambu_quasi-particles_1960,anderson_plasmons_1963,higgs_broken_1964}.
In addition, in a superconductor one can excite fluctuations of the amplitude of the superconducting order parameter. The amplitude mode is a gapped mode with minimal excitation energy $2\Delta$, i.e., equal to the superconducting gap and is called the Higgs mode in analogy to the Higgs boson in particle physics~\cite{higgs_broken_1964}. Since the Nambu-Goldstone modes are shifted to the plasma frequency, the Higgs mode is the lowest-energy collective excitation of the superconducting order parameter and, therefore, stable against a decay into the phase mode~\cite{volkov_collisionless_1974}.

An experimental detection of the Higgs mode is challenging for a number of reasons. First of all, it is a charge-neutral mode that does not couple directly to electromagnetic fields. In addition, for typical BCS superconductors the energy of the Higgs mode is in the terahertz (THz) regime where until recently there was a lack of suitable sources to excite the system. Finally, the Higgs mode energy of $2\Delta$ equals the threshold for single-particle excitations which makes it difficult to excite the Higgs mode without exciting quasiparticles at the same time.

Experimentally, the Higgs mode has been observed for the first time by Raman scattering in materials that are both superconducting and show a charge density wave~\cite{sooryakumar_raman_1980,sooryakumar_raman_1981}. 
Recent advances in the field of THz radiation have allowed for the excitation of the Higgs mode by monocycle THz pump pulses and its subsequent observation via the transient oscillation of the transmitted THz probe radiation~\cite{matsunaga_higgs_2013}. The experimental results have been explained in terms of the dynamics of Anderson's pseudospin in a two-dimensional BCS model~\cite{chou_twisting_2017} and within a gauge-invariant microscopic kinetic theory of superconductivity~\cite{yang_gauge-invariant_2019}. In addition, the Higgs mode excited by THz pulses has also been probed by third-harmonic generation~\cite{matsunaga_light-induced_2014} which arises due to the nonlinear coupling to electromagnetic fields~\cite{tsuji_theory_2015,schwarz_theory_2020}
The Higgs mode has also been observed by THz spectroscopy of thin, disordered superconducting films close to an insulator-superconductor quantum phase transition where it manifests itself as an excess absorption at energies below the superconducting gap~\cite{sherman_higgs_2015}.
Recently, it has been demonstrated experimentally that in the presence of supercurrents the Higgs mode becomes infrared active and gives rise to a sharp resonant peak in the optical conductivity at the Higgs frequency~\cite{nakamura_infrared_2019}.

Additional theoretical works have studied the Higgs mode in unconventional superconductors~\cite{schwarz_classification_2020} as well as the interplay of Higgs and Leggett modes in multi-band superconductors~\cite{krull_coupling_2016,giorgianni_leggett_2019}. Furthermore, the occurrence of Higgs mode in superconductor-normal metal junctions~\cite{vadimov_higgs_2019} and its signatures in transport properties have been analyzed~\cite{tang_signatures_2020,silaev_spin_2020}. Recent reviews on Higgs physics in superconductors can be found in Refs. ~\cite{pekker_amplitude/higgs_2015,shimano_higgs_2020}.

In this paper, we investigate the dynamics of superconducting correlations in a time-dependently driven superconductor-quantum dot hybrid structure. Superconductor-quantum dot heterostructures have been studied intensively both from a theoretical as well as from an experimental perspective, see Refs.~\cite{martin-rodero_josephson_2011,de_franceschi_hybrid_2010} for recent reviews. They exhibit an exciting playground to study the interplay between the superconducting proximity effect, strong Coulomb interactions, and transport situations far from equilibrium. Furthermore, they provide a high degree of tunability by applying, e.g., magnetic fields or gate voltages.  Studying the pair amplitude dynamics of a quantum dot allows us to analyze the coherent dynamics of a single Cooper pair rather than the collective dynamics of all Cooper pairs which gives rise to the order parameter dynamics in bulk superconductors. Furthermore, in a quantum dot system, the Cooper pair dynamics can be investigated under different forms of external driving such as parameter quenches and periodic driving in situations far away from equilibrium. 
As we will demonstrate below, a nonequilibrium situation induced, e.g., by a temperature bias between the superconducting reservoirs is crucial in establishing a coherent pair amplitude dynamics on the quantum dot even in the weak-coupling limit.

At this point, we would like to point out that there are some fundamental differences between the pair amplitude dynamics of the quantum dot and the Higgs mode in bulk superconductors. First of all, in the bulk case, the central quantity of interest is the superconducting order parameter $\Delta$ which is the macroscopic wave function of the whole superconducting condensate. In contrast, the pair amplitude of the quantum dot describes a single Cooper pair on the dot which is coupled to macroscopic condensate in the superconducting reservoirs. Second, the equilibrium value of the bulk order parameter is determined by the maximization of the free energy while the stationary, nonequilibrium value of the dot's pair amplitude is determined by the generalized master equation. While the dynamics of both the bulk order parameter as well as the dot's pair amplitude are governed by a Bloch-type equation for the Anderson pseudospin, the precession frequency of the pseudospin degree of freedom is given by the superconducting gap $2|\Delta|$ in the bulk case while it is determined by an effective exchange field $\vec B_\text{ex}$ in the quantum dot case which depends on the tunnel coupling, the Coulomb interaction, the superconducting order parameter of the electrodes and the dot level detuning. Furthermore, the dynamics of the bulk order parameter does not couple directly to external electromagnetic fields while the dynamics of the pair amplitude is directly connected to the gate voltage applied to the quantum dot. 
Since the Higgs mode is difficult to excite in bulk superconductors, the order parameter oscillates by less than ten percent in typical present experiments. In contrast, the quantum-dot system allows for much larger oscillations of the superconducting pair amplitude. 
Finally, in the bulk case the Higgs mode is the only low-energy excitation of the order parameter while the Nambu-Goldstone mode is shifted to the plasma frequency by the Anderson-Higgs mechanism. In contrast, for the pair amplitude on the dot one can excite both amplitude and phases modes at low energies.

The paper is organized as follows. In Sec.~\ref{sec:model}, we present our theoretical model of a superconductor-quantum dot hybrid structure. We discuss the real-time diagrammatic approach used to analyze the system in Sec.~\ref{sec:method}. The results for the pair amplitude dynamics on the quantum dot after a quench and under periodic driving are analyzed in Sec.~\ref{ssec:quench} and Sec.~\ref{ssec:periodic}, respectively. We conclude by comparing the pair amplitude dynamics with the order parameter dynamics of bulk superconductors in Sec.~\ref{sec:discussion}.

\section{\label{sec:model}Model}
\begin{figure}
    \includegraphics[width=\columnwidth]{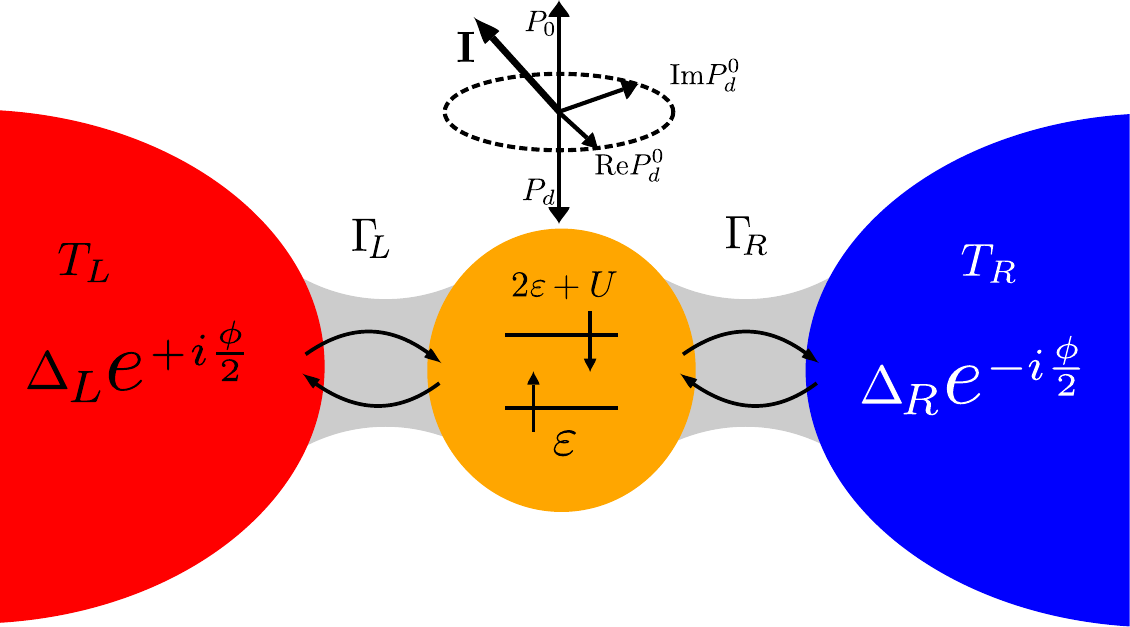}
	\caption{\label{fig:model}Schematic sketch of a system. A single-level quantum dot is tunnel coupled to two superconducting reservoirs $\eta=\text{L,R}$ at different temperatures $T_\eta$. The system is subject to time-dependent driving via modulations of the level position or a superconducting phase difference that changes with time. The driving gives rise to a nontrivial dynamics of the superconducting pair amplitude induced on the dot via the proximity effect which is characterized by the pseudospin $\vec I$ that describe coherent superpositions of the empty and doubly occupied dot state.}
\end{figure}

We consider a single-level quantum dot weakly tunnel coupled to two superconducting electrodes $\eta=\text{L,R}$, see Fig.~\ref{fig:model}. Both superconductors are kept at the same electrochemical potential but can have different temperatures $T_\eta$, thus driving the system into a stationary nonequilibrium state. In addition, the system is subject to a time-dependent driving of either the superconducting phase difference $\phi(t)$ or the level position of the quantum dot $\varepsilon(t)$ that can be tuned by an applied gate voltage.
The setup is described by the total Hamiltonian
\begin{equation}\label{eq:Hamiltonian}
	H=\sum_\eta H_\eta+H_\text{dot}+H_\text{tun}.
\end{equation}
The first term describes the two superconducting electrodes in terms of the mean-field BCS Hamiltonian 
\begin{equation}
	H_\eta=\sum_{\vec k\sigma}\varepsilon_{\eta \vec k}a_{\eta\vec k\sigma}^\dagger a_{\eta\vec k\sigma}+\Delta_\eta e^{i\phi_\eta}\sum_{\vec k}a_{\eta \vec k\up}a_{\eta -\vec k\down}+\text{H.c.},
\end{equation}
where the first term corresponds to the kinetic energy of electrons in lead $\eta$ with spin $\sigma$ and momentum $\vec k$. The second term describes the superconducting pairing. The superconducting order parameter is characterized by its phase $\phi_\text{L}(t)=-\phi_\text{R}(t)=\phi(t)/2$ and its absolute value $\Delta_\eta$. We assume both superconductors to have the same critical temperature $T_c$ and, therefore, to have the same absolute value of the order parameter at zero temperature, $\Delta_0=1.764\kBT_c$. The temperature dependence of $\Delta_\eta$ follows from a self-consistency equation that can be solved only numerically. However, the temperature dependence can be approximated with an error of less than 2 percent as
\begin{equation}
	\Delta_\eta(T_\eta)=\Delta_0\tanh\left(1.74\sqrt{\frac{T_c}{T_\eta}-1}\right)
\end{equation}
in the whole temperature range from $T_\eta=0$ to $T_\eta=T_{c}$.
We assume the density of states of the leads in the normal state $\rho_\eta^\text{N}$ to be independent of energy which is a reasonable approximation for the energy scales involved in our considerations. The density of states in the superconducting states normalized to $\rho_\eta^\text{N}$ is then given by the standard BCS expression
\begin{equation}
	\rho_\eta^\text{BCS}(E)=\frac{|E|\Theta(|E|-|\Delta_\eta|)}{\sqrt{E^2-\Delta_\eta^2}}.
\end{equation}

The second term in Eq.\eqref{eq:Hamiltonian} describes the quantum dot in terms of a single, spin-degenerate level with time-dependent level position $\varepsilon(t)$ as 
\begin{equation}
	H_\text{dot}=\sum_\sigma \varepsilon(t) c_\sigma^\dagger c_\sigma +Uc_\up^\dagger c_\up c_\down^\dagger c_\down,
\end{equation}
where the second term denotes the Coulomb energy $U$ that is necessary to occupy the quantum dot with two electrons at the same time. As we will detail below, the superconducting proximity effect which gives rise to a finite superconducting pair amplitude on the dot has a sizeable effect only if the empty and doubly occupied state are quasi-degenerate. This happens close to the particle-hole symmetric point $\varepsilon=-U/2$. We characterize deviations from this point by the time-dependent detuning $\delta(t)=2\varepsilon(t)+U$.

The coupling between the dot and the leads is given by
\begin{equation}
	H_\text{tun}=\sum_{\eta \vec k\sigma}t_\eta a_{\eta\vec k\sigma}^\dagger c_\sigma+\text{H.c.}
\end{equation}
where we assume the tunnel matrix elements $t_\eta$ to be independent of spin and momentum. They are related to the tunnel coupling strengths $\Gamma_\eta=2\pi|t_\eta|^2\rho_\eta^\text{N}$. The total tunnel coupling is given by $\Gamma=\Gamma_\text{L}+\Gamma_\text{R}$.

\section{\label{sec:method}Real-time diagrammatic transport theory}
In order to describe the dynamics of the superconducting pair amplitude induced on the quantum dot by the coupling to the superconducting reservoirs, we make use of a real-time diagrammatic approach~\cite{konig_zero-bias_1996,konig_resonant_1996,konig_quantum_1999} in its extension to superconducting leads~\cite{governale_real-time_2008,governale_erratum:_2008,kamp_phase-dependent_2019}. It allows us to describe arbitrary nonequilibrium situations, to take into account the Coulomb interaction on the quantum dot exactly and to perform a systematic perturbation expansion in the tunnel couplings. The real-time diagrammatic approach is based on the idea to integrate out the noninteracting reservoir degrees of freedom and to describe the quantum dot degrees of freedom in terms of a reduced density matrix $\rho_\text{red}$ with matrix elements $P^{\chi_1}_{\chi_2}=\bra{\chi_1}\rho_\text{red}\ket{\chi_2}$. Here, $\chi_{1,2}$ denote the eigenstates of the quantum-dot Hamiltonian, i.e. the empty dot $\ket{0}$, the dot occupied with a spin-up electron $\ket{\up}$ or a spin-down electron $\ket{\down}$ and the doubly occupied dot $\ket{d}$. We remark that the formulation of the real-time diagrammatics in Ref.~\cite{governale_real-time_2008,governale_erratum:_2008} also explicitly accounts for the number of Cooper pairs in the leads. This particle-number conserving formulation is required to properly describe situations in which a bias voltage is applied between superconducting leads. Since we focus on a situation where all superconducting leads are at the same electrochemical potential, we can simplify the description and drop the number of Cooper pairs in the leads.
The time evolution of the reduced density matrix is given by a generalized master equation of the form
\begin{widetext}
\begin{equation}\label{eq:GME}
	\frac{dP^{\chi_1}_{\chi_2}(t)}{dt}=-i(E_{\chi_1}-E_{\chi_2})P^{\chi_1}_{\chi_2}(t)+\sum_{\chi'_1,\chi'_2}\int_{-\infty}^t dt'W^{\chi_1\chi'_1}_{\chi_2\chi'_2}(t,t')P^{\chi'_1}_{\chi'_2}(t')
\end{equation}
\end{widetext}
The first term on the right-hand side describes the coherent evolution of the quantum dot system. The second term arises due to the dissipative coupling to the superconducting reservoirs. The generalized transition rates $W^{\chi_1\chi'_1}_{\chi_2\chi'_2}$ are evaluated as irreducible self-energy blocks of the quantum dot propagator on the Keldysh contour. In the following, we will take into account tunneling processes up to first order in the tunnel coupling only. This accounts for both, normal tunneling processes as well as Andreev processes. The latter give rise to the superconducting proximity effect on the quantum dot and induce a finite superconducting pair amplitude on the dot. We emphasize though that there is no Josephson current through the quantum dot due to first-order processes. The latter require a coherent charge transfer between the two superconducting leads and can, therefore, occur only in second and higher order processes~\cite{kamp_phase-dependent_2019}.

The time-dependent driving of the system affects the dynamics of reduced density matrix in two ways. First, it gives rise to a dependence of the generalized transition rates on driven parameters. Second, it introduces memory effects, i.e., density matrix elements at time $t$ depend on density matrix elements at earlier times $t'$. A systematic treatment of these non-Markovian effects has been developed in the framework of the real-time diagrammatic approach in Refs.~\cite{splettstoesser_adiabatic_2006,cavaliere_nonadiabatic_2009}. In particular, non-Markovian effects also modify the form of the generalized master equation by giving rise to an additional inhomogeneity~\cite{flindt_counting_2008,marcos_non-markovian_2011,stegmann_real-time_2020}.
In the following, we will take into account only the Markovian contributions to the dynamics. This is motivated by the fact that we expect the order parameter dynamics that we are interested in to take place on time scales larger than the inverse tunnel coupling whereas memory effects are relevant for times much shorter than the inverse tunnel couplings.
The form of the generalized master equation~\eqref{eq:GME} implies that coherent superpositions between two dot states are only possible if the energy splitting between the states is comparable to the generalized transition rates, i.e., if it is of the order of the tunnel coupling. Hence, for the superconductor-quantum dot system, coherent superpositions of the empty and doubly occupied state can occur only for $\delta(t)\sim\Gamma$. In addition, we require the Coulomb energy to be larger than the superconducting gaps, $U>2\Delta_\eta$, because otherwise no tunneling of quasiparticles is possible to lowest order in $\Gamma$.

We can cast the master equation into a physically more intuitive form by introducing the probabilities to find the dot occupied with an even or odd number of electrons 
\begin{equation}
	\vec P=\left(\begin{array}{c} P_\text{e}\\P_\text{o}\end{array}\right)=\left(\begin{array}{c}P_0+P_d\\P_\up+P_\down\end{array}\right).
\end{equation}
Furthermore, we introduce a pseudospin degree of freedom involving the empty and doubly occupied dot state in analogy to Anderson's pseudospin as
\begin{equation}
	\vec I=\left(\begin{array}{c} I_x\\I_y\\I_z \end{array}\right)=\left(\begin{array}{c}\re P^d_0 \\ \im P^d_0 \\ \frac{P_0-P_d}{2}\end{array}\right).
\end{equation}
With the above definitions, we can decompose the generalized master equation into one set that describes the time evolution of the occupation probabilities
\begin{equation}\label{eq:MEP}
	\frac{d\vec P}{dt}=\sum_\eta\left[\left(\begin{array}{cc}-Z^-_\eta & Z^+_\eta \\ Z^-_\eta & -Z^+_\eta \end{array}\right)\vec P+4X^-_\eta\left(\begin{array}{c} 1 \\ -1 \end{array}\right) \vec I\cdot\vec n_\eta\right].
\end{equation}
The dot occupation can change due to tunneling in and out of electrons with rates
\begin{equation}
	Z^\pm_\eta=\frac{2\Gamma_\eta}{\hbar}\rho^\text{BCS}_\eta(U/2)f_\eta(\pm U/2)
\end{equation}
as described by the first term. Here, $f_\eta(\omega)=[\exp(\omega/\kBT_\eta)+1]^{-1}$ denotes the Fermi function of lead $\eta$. In addition, the dot occupation is also influenced by the pseudospin accumulation via the second term where the rates
\begin{equation}
	X^\pm_\eta=\pm\frac{2\Gamma_\eta}{\hbar}\frac{\Delta_\eta}{U}\rho^\text{BCS}_\eta(U/2)f_\eta(\pm U/2)
\end{equation}
are due to Andreev processes and $\vec n_\eta=(\cos\phi_\eta(t),\sin\phi_\eta(t),0)$ denotes a unit vector that characterizes the phase of the superconducting order parameter in the leads.
We remark that the rates $Z^\pm_\eta$ and $X^\pm_\eta$ formally diverge for $U/2=\Delta_\eta$ due to the divergence of the BCS density of states at the gap edge. In a real system, the BCS density of states is smeared out, thus turning the divergence into a pronounced peak and rendering the transition rates finite at $U/2=\Delta_\eta$.
A second set of equations describes the time evolution of the pseudospin as
\begin{equation}\label{eq:MEI}
	\frac{d\vec I}{dt}=\left(\frac{d\vec I}{dt}\right)_\text{acc}-\frac{\vec I}{\tau_\text{rel}}+\vec B_\text{ex}\times\vec I.
\end{equation}
The first term on the right-hand side accounts for the accumulation of pseudospin on the dot due to the tunneling in and out of electrons
\begin{equation}
	\left(\frac{d\vec I}{dt}\right)_\text{acc}=\sum_\eta \left(X^-_\eta P_\text{e}+X^+_\eta P_\text{o}\right)\vec n_\eta.
\end{equation}
The second term describes a relaxation of the pseudospin on a timescale $1/\tau_\text{rel}=\sum_\eta Z^-_\eta$ which is also caused by the tunneling of electrons. Finally, the third term gives rise to a coherent precession of the pseudospin in an effective exchange field given by 
\begin{equation}
	\vec B_\text{ex}=\sum_\eta B_\eta\vec n_\eta+\delta(t) \vec e_z,
\end{equation}
where
\begin{equation}
	B_\eta=\frac{2\Gamma_\eta}{\pi\hbar}\int' d\omega \rho_\eta^\text{BCS}(\omega)\frac{f_\eta(\omega)}{\omega+U/2}\sign\omega.
\end{equation}
The exchange field arises from virtual Andreev tunneling processes between the dot and the superconductors which renormalize the excitation energies of the empty and doubly occupied dot state relative to each other. Interestingly, the level renormalization which is of the order of the tunnel coupling $\Gamma$ impacts the pseudospin dynamics already in sequential tunneling because the dwell time of electrons on the dot scales as $\Gamma^{-1}$ such that the precession angle is of the order $\mathcal O(\Gamma^0)$. In addition, the exchange field has a contribution along the $z$ axis which arises from the breaking of particle-hole symmetry by the detuning $\delta$.

We remark that the fact that the generalized master equations can be formulated in the coordinate-free version shown above is linked to gauge invariance, i.e. to the fact that the phases of all superconducting pair amplitudes can be changed by the same amount without changing the underlying physics.

A central quantity of interest in the following is the time-dependent, proximity-induced superconducting pair amplitude on the quantum dot. It is given by
\begin{equation}
	\mathcal F=\langle c_\down c_\up \rangle,
\end{equation}
such that its absolute value $|\mathcal F|$ can be expressed in terms of the pseudospin as
\begin{equation}
	|\mathcal F|=\sqrt{I_x^2+I_y^2},
\end{equation}
while its phase $\Phi$ is given by
\begin{equation}
	\Phi=\arctan \frac{I_y}{I_x}.
\end{equation}
A time-dependent modulation of the absolute value $|\mathcal F|$ can be considered as an analogue of the Higgs mode in bulk superconductors while a modulation of $\Phi$ represents the analogue of Nambu-Goldstone mode. We remark that the phase $\Phi$ is itself not a physical observable but should always be considered relative to the phases of the two superconducting reservoirs.

\section{\label{sec:results}Results}
In the following, we are going to discuss the dynamics of the superconducting pair amplitude induced on the quantum dot via the proximity effect. We will start our analysis by considering the dynamics after a quench in Sec.~\ref{ssec:quench} and then turn to the dynamics in the case of a periodic driving in Sec.~\ref{ssec:periodic}.

\subsection{\label{ssec:quench}Quench dynamics}
We analyze the pair amplitude dynamics after a quench of the system parameters. To this end, we focus on the situation where the dot is coupled to a single superconductor because this scenario contains already all essential features of the dynamics but can be tackled fully analytically at the same time. To keep our notation as simple as possible, we omit the lead index $\eta$ and choose the phase $\phi=0$ in the remaining discussion of the quench dynamics. Furthermore, we focus on the particle-hole symmetric point $\delta=0$ to obtain compact analytical expressions. We remark that away from $\delta=0$ there are no qualitatively new features in the relaxation dynamics.

We consider a situation where the dot is prepared in an arbitrary initial state. At time $t=0$, the dot is coupled to the superconducting reservoir. The subsequent relaxation dynamics is given by the solution of the generalized master equation as
\begin{widetext}
	\begin{eqnarray}
		P_\text{e}(t)&=&f(U/2)+\frac{e^{\gamma_{\text{ch}3}t}+e^{\gamma_\text{p}t}}{2}\left(P_e^{(0)}-f(U/2)\right)+\frac{e^{\gamma_{\text{ch}3}t}-e^{\gamma_\text{p}t}}{2\Omega}\left[Z^+\left(P_e^{(0)}-f(U/2)\right)-8X^-I_x^{(0)}\right],\\
		P_\text{o}(t)&=&1-P_\text{e}(t),\\
		S_z(t)&=&S_z^{(0)}e^{\gamma_\text{spin}t},\\
		I_x(t)&=&\frac{e^{\gamma_{\text{ch}3}t}+e^{\gamma_\text{p}t}}{2}I_x^{(0)}+\frac{e^{\gamma_{\text{ch}3}t}-e^{\gamma_\text{p}t}}{2\Omega}\left[Z^-I_x^{(0)}+2\left(X^+-X^-\right)\left(f(U/2)-P_e^{(0)}\right)\right],\\
		I_y(t)&=&e^{\frac{\gamma_{\text{ch}1}+\gamma_{\text{ch}2}}{2}t}\left(I_y^{(0)}\cos Bt-I_z^{(0)}\sin Bt\right),\\
		I_z(t)&=&e^{\frac{\gamma_{\text{ch}1}+\gamma_{\text{ch}2}}{2}t}\left(I_y^{(0)}\sin Bt+I_z^{(0)}\cos Bt\right),
	\end{eqnarray}
\end{widetext}
where density matrix elements with a superscript such as $P_\text{e}^{(0)}$ denote the values at the initial time $t=0$.

The time evolution of the occupation probabilities is governed by an exponential decay towards the equilibrium occupation $P_\text{e}=1-P_\text{o}=f(U/2)$ with two rates, $\gamma_{\text{ch}3}=-Z^--(Z^+-\Omega)/2$ and $\gamma_\text{p}=-Z^--(Z^++\Omega)/2$ where $\Omega=\sqrt{16X^-(X^--X^+)+(Z^+)^2}$. The decay is driven by the nonequilibrium occupation of the dot but is also influenced by a finite pseudospin accumulation along the $x$ axis in the initial state.
Any (real) spin accumulation along the $z$-axis on the dot follows a simple exponential decay with rate $\gamma_\text{spin}=-Z^+$.
The $x$-component of the pseudospin shows an exponential decay of the initial pseudospin accumulation with rates $\gamma_\text{p}$ and $\gamma_{\text{ch}3}$ but is furthermore affected by a nonequilibrium dot occupation on the same time scale. As we will demonstrate below, the latter term can even give rise to an initial increase of the pseudospin right after the quench.
Finally, the $y$- and $z$-component of the pseudospin both decay with the rate $(\gamma_{\text{ch}1}+\gamma_{\text{ch}2})/2$ where $\gamma_{\text{ch}1,2}=-Z^-\pm iB$ as expected from the master equation~\eqref{eq:MEI}. In addition, the two pseudospin components show an oscillatory behavior with a frequency given by the exchange field $B$.

\begin{figure}
	\includegraphics[width=\columnwidth]{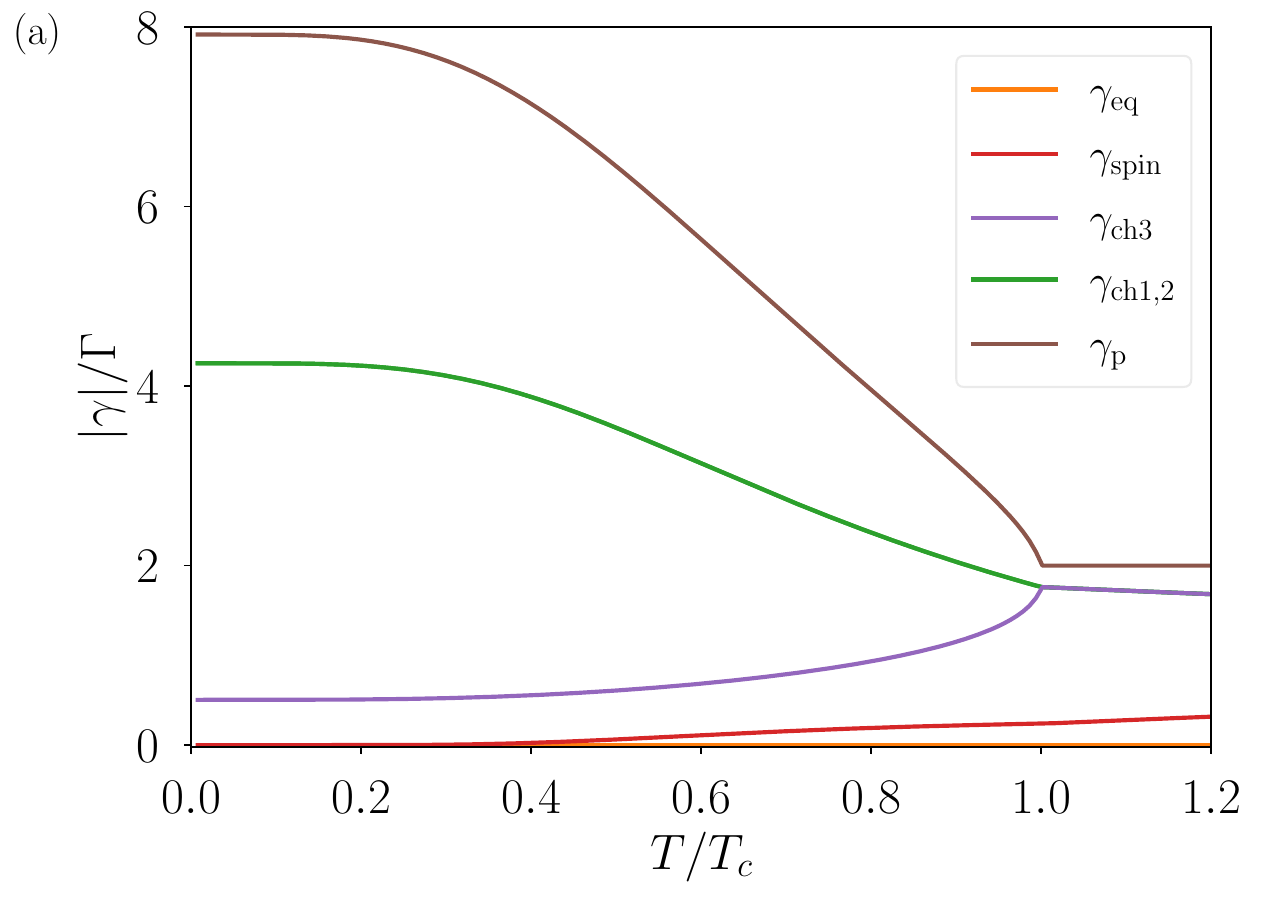}
	\includegraphics[width=\columnwidth]{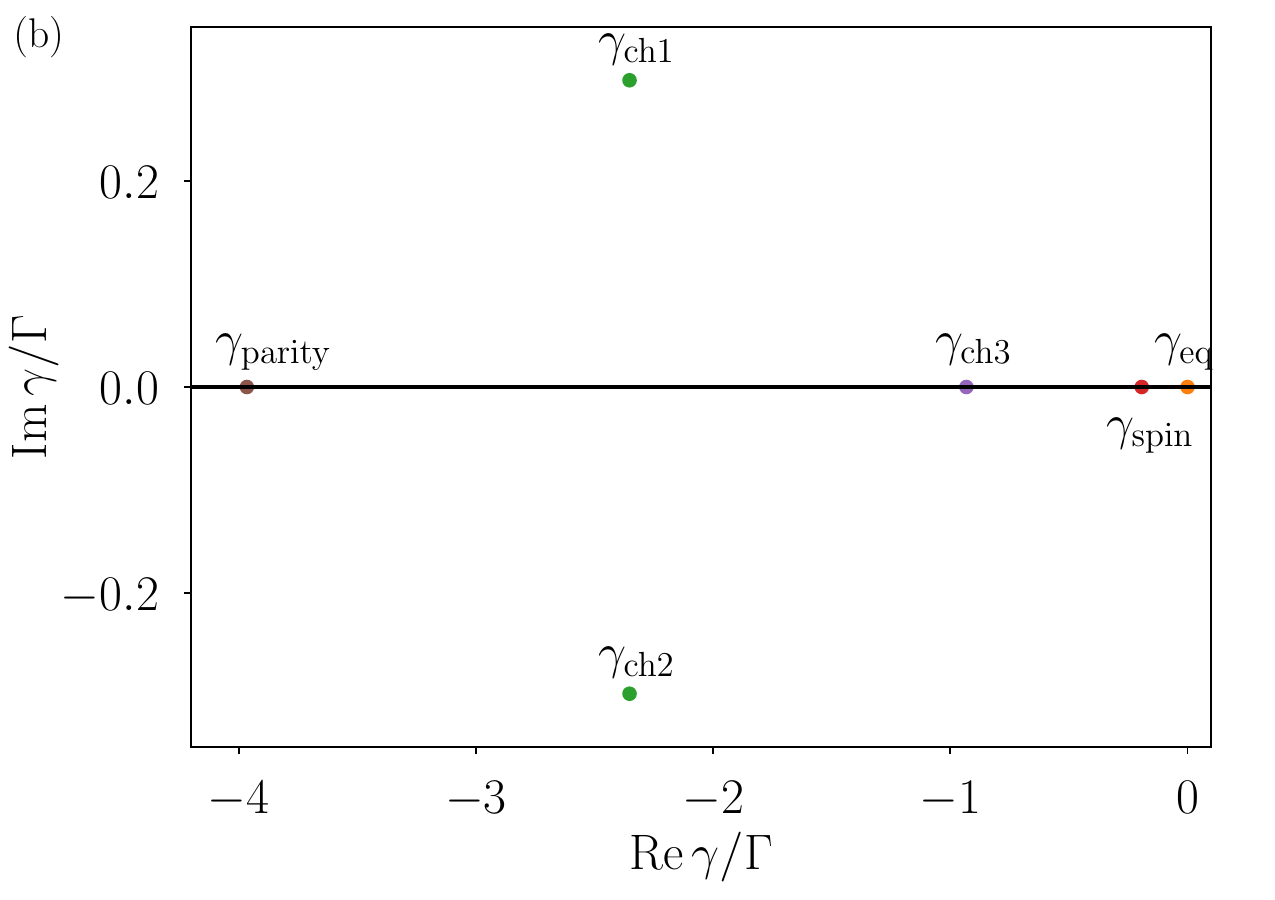}
	\caption{\label{fig:eigenvalues}(a) Absolute value of the relaxation rates $\gamma_i$ as function of temperature. (b) Real and imaginary part of the relaxation rates $\gamma_i$. Parameters are $T=0.8T_c$, $U=4\kBT_c$, and $\delta=0$.}
\end{figure}

The quantities $\gamma_{\text{ch}1-3}$, $\gamma_\text{p}$, and $\gamma_\text{spin}$ together with $\gamma_\text{eq}=0$ are the eigenvalues of the transition rate matrix. Their temperature dependence is shown in Fig.~\ref{fig:eigenvalues}. Above the critical temperature $T_c$ of the superconducting lead, the three eigenvalues $\gamma_{\text{ch}1-3}$ are identical. The resulting four different eigenvalues all have an intuitive physical interpretation~\cite{splettstoesser_charge_2010,contreras-pulido_time_2012,saptsov_fermionic_2012}. The eigenvalue $\gamma_\text{eq}=0$ which is independent of temperature is related to the stationary state of the system. The eigenvalue $\gamma_\text{spin}$ describes the decay of spin accumulation. The decay of charge on the dot is governed by $\gamma_\text{{ch}1-3}=-Z^-$ while the decay of the occupation parity is determined by $\gamma_\text{p}=-2\Gamma/\hbar$. 

Below the critical temperature, the three eigenvalues $\gamma_{\text{ch}1-3}$ split into one pair of complex eigenvalues and a third, real one, cf. Fig.~\ref{fig:eigenvalues}(b) and the corresponding analytical expressions above. As can be seen from the time evolution of the density matrix elements after the quench, the eigenvalues $\gamma_{\text{ch}1-3}$ and $\gamma_\text{p}$ no longer describe the decay of charge and parity, respectively but rather account for the decay of linear combinations of charge, parity and pseudospin.
As the temperature is lowered the spin decay rate decreases because it is exponentially suppressed by $U/\kBT$. At the same time, $\gamma_\text{p}$ increases significantly with decreasing temperature because the superconducting density of states is enhanced by the factor $\rho^\text{BCS}(U/2)$ compared to the normal-conducting case.

\begin{figure}
	\includegraphics[width=\columnwidth]{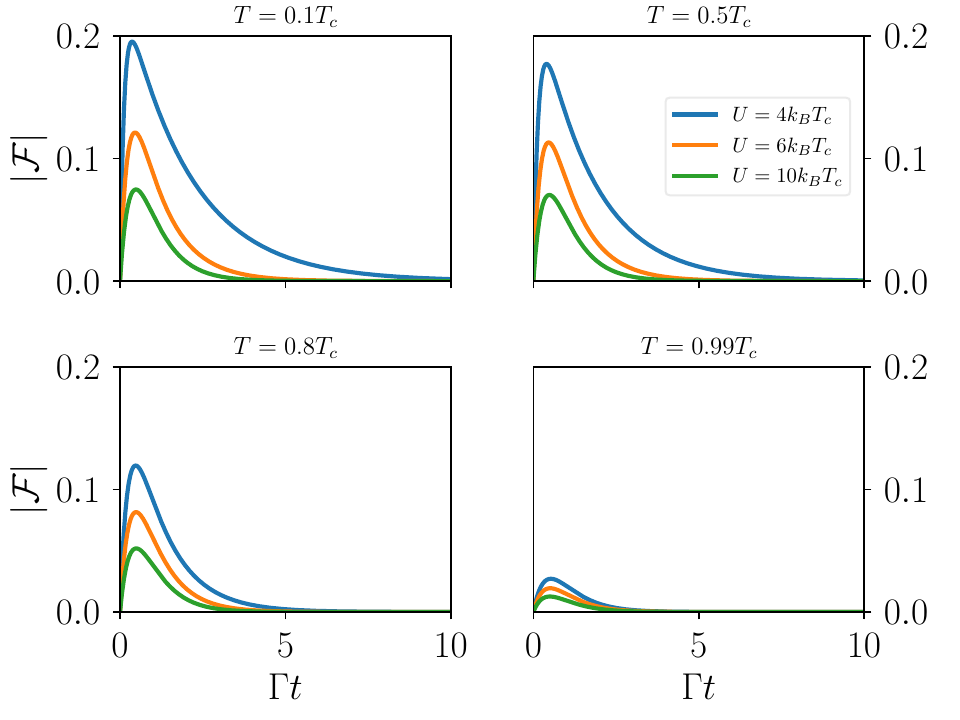}
	\caption{\label{fig:Fquench}Time evolution of the absolute value of the superconducting pair amplitude $|\mathcal F|$ after a quench for different temperatures and different Coulomb interactions. The detuning is chosen as $\delta=0$.}
\end{figure}

We illustrate our general consideration of the dot dynamics after a quench with the concrete example of a quantum dot that is prepared in the empty state, i.e. with $P_\text{e}^{(0)}=1$ and $I_z^{(0)}=1/2$ and all other density matrix elements zero. The resulting time-dependence of the absolute value of the superconducting pair amplitude on the quantum dot is given by
\begin{equation}\label{eq:Fquench}
	|\mathcal F|=e^{-Z^-t}\sqrt{\frac{4e^{-Z^+t}\left(X^-\right)^2\sinh^2\Omega t}{\Omega^2}+\frac{1}{4}\sin^2 Bt}.
\end{equation}
It is depicted in Fig.~\ref{fig:Fquench} for different temperatures and different strengths of the Coulomb interaction.
At short times, the pair amplitude grows linearly in time due to real and virtual tunneling processes between the dot and the lead. It reaches a maximum at times $\Gamma t\sim 1$. 
The maximal value is suppressed by large Coulomb interactions as these are detrimental to the proximity effect. Furthermore, we find that the maximal pair amplitude increases as the temperature is lowered because the rate $X^-$ that governs the pseudospin accumulation grows as temperature is decreasing. For longer times, the pair amplitude decays exponentially towards zero with a rate given by $\gamma_{\text{ch}3}$.
In addition to the exponential decay, the pair amplitude also shows an additional oscillatory time dependence on a time scale give by the inverse exchange field as can be seen in Eq.~\eqref{eq:Fquench}.

Both, the precession and the exponential decay occur with rates and frequencies of the order of the tunnel coupling $\Gamma$. However, our numerical analysis reveals that the exchange field is about a factor of ten smaller than the decay rates. This is because the decay rate gets enhanced by the BCS density of states of the lead while the exchange field which contains contributions from electrons at all energies is suppressed by the occurrence of the superconducting gap. As a result, the precessional dynamics of the pseudospin that can be viewed as an analogue of the Higgs mode in bulk superconductors is hardly visible after a quench.

\begin{figure}
	\includegraphics[width=\columnwidth]{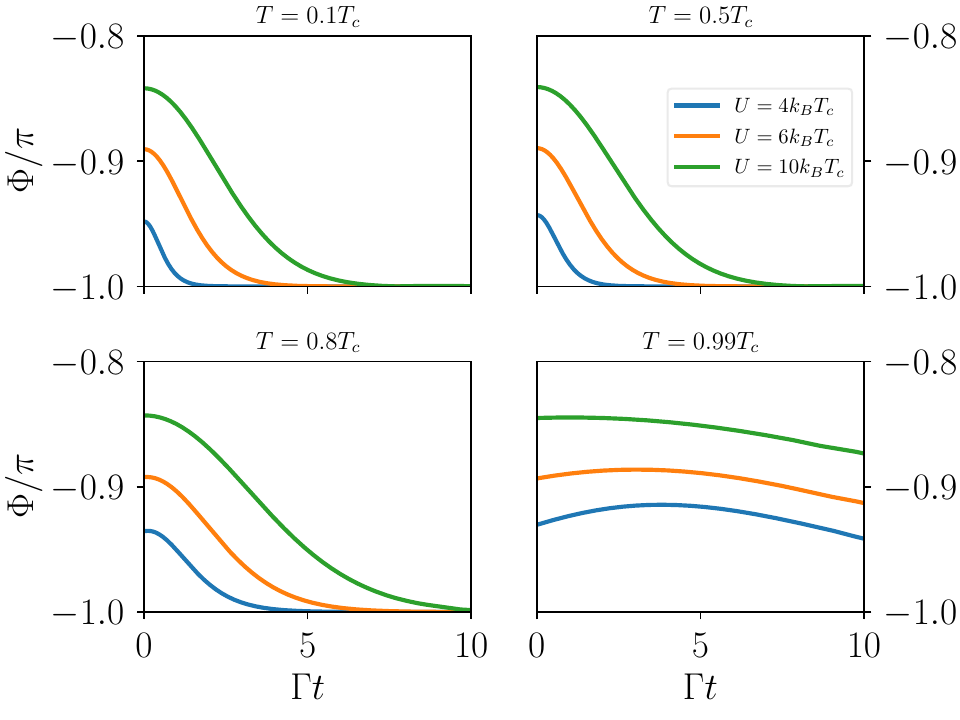}
	\caption{\label{fig:Phiquench}Time evolution of the phase of the superconducting pair amplitude $\Phi$ after a quench for different temperatures and different Coulomb interactions. The detuning is chosen as $\delta=0$.}
\end{figure}

The phase of the pair amplitude after the quench is given by
\begin{equation}\label{eq:Phiquench}
	\tan\Phi=-\frac{\Omega}{4}\frac{e^{Z^+t/2}\sin Bt}{X^-\sinh\Omega t}.
\end{equation}
Its time evolution is shown in Fig.~\ref{fig:Phiquench}. As the quench takes place, the phase jumps to a finite value and subsequently decays to $-\pi$. The decay becomes slower as the Coulomb interaction is increased because the BCS density of states is smaller at larger energies. In addition, the decay becomes slower as the temperature approaches the critical temperature of the superconducting lead. This is a consequence of critical slowing down. The time scale of the decay scales as $|\tau|^{-1}$ where $\tau=(T-T_c)/T_c$ denotes the reduced temperature. This behavior corresponds to a critical exponent of one in agreement with the expectation of mean-field theory. We remark that the scaling behavior occurs only for temperatures extremely close to the phase transitions where deviations from the BCS mean-field descriptions are expected to become relevant. Furthermore, as the induced pair amplitude goes to zero as the critical temperature is approached, an experimental observation of the critical slowing down seems to be experimentally very challenging.

\subsection{\label{ssec:periodic}Periodic driving}
As we have just discussed, the dynamics of the system after a quench is dominated by an exponential relaxation towards equilibrium because damping occurs on shorter time scales than coherent oscillations. To overcome this issue, we are going to analyze the system dynamics under a continuous, periodic driving in the following. To this end, we consider a situation where the dot is coupled to two superconducting reservoirs. A temperature bias between the leads drives the dot into a static nonequilibrium state with a finite pair amplitude. The time-dependent, periodic driving of either the phase difference $\phi(t)$ or the dot level detuning $\delta(t)$ with frequency $\omega$ then gives rise to a nontrivial dynamics of the pair amplitude. We are going to study the dynamics in three different parameter regimes. First, we will consider the case of adiabatic driving in Sec.~\ref{sssec:adiabatic} where the dynamics can be understood from the properties of the stationary state. Next, we turn to the case of fast driving in Sec.~\ref{sssec:fast}. Finally, we will address the intermediate regime in Sec.~\ref{sssec:intermediate}.

\subsubsection{\label{sssec:adiabatic}Adiabatic driving}
\begin{figure}
	\includegraphics[width=\columnwidth]{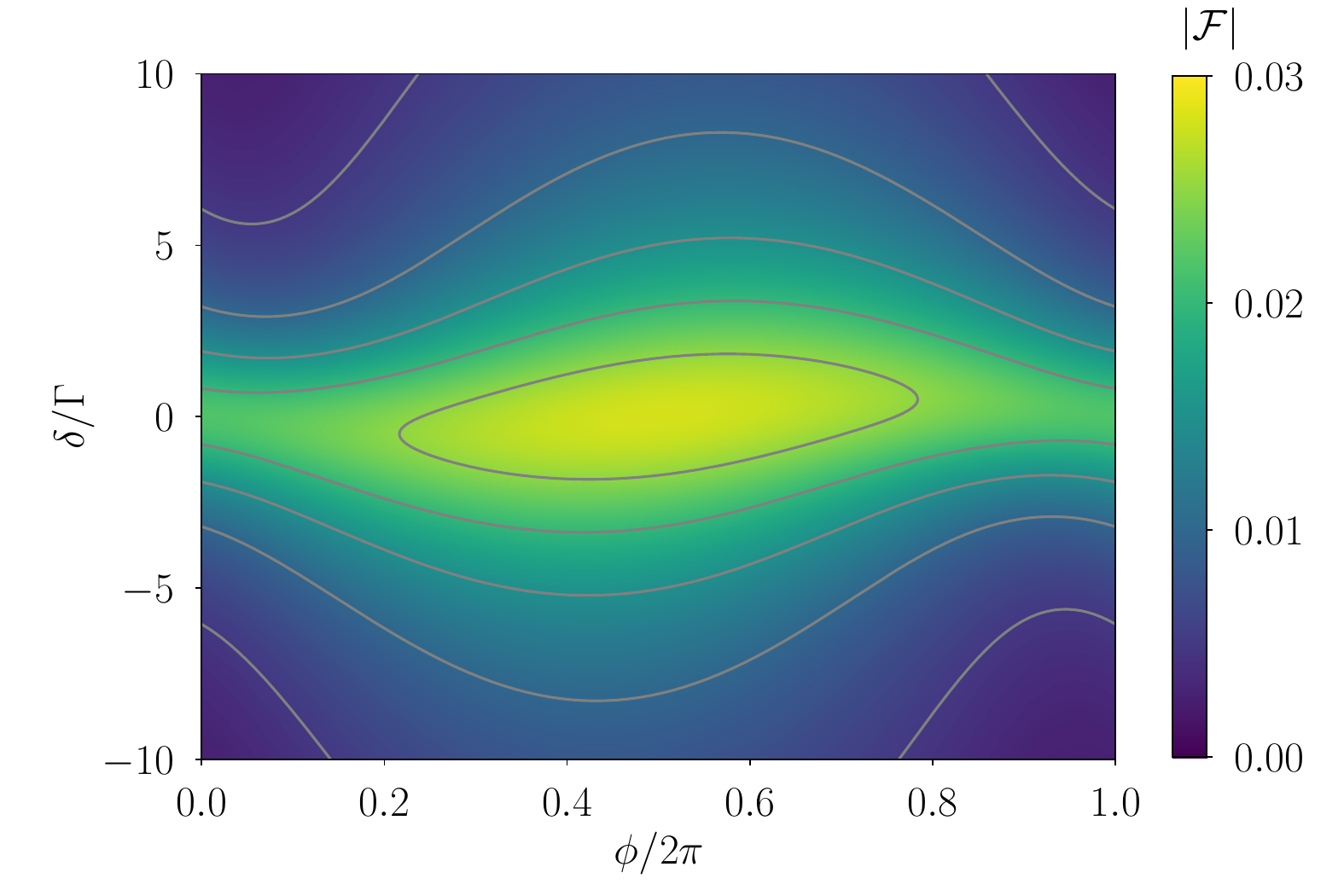}
	\includegraphics[width=\columnwidth]{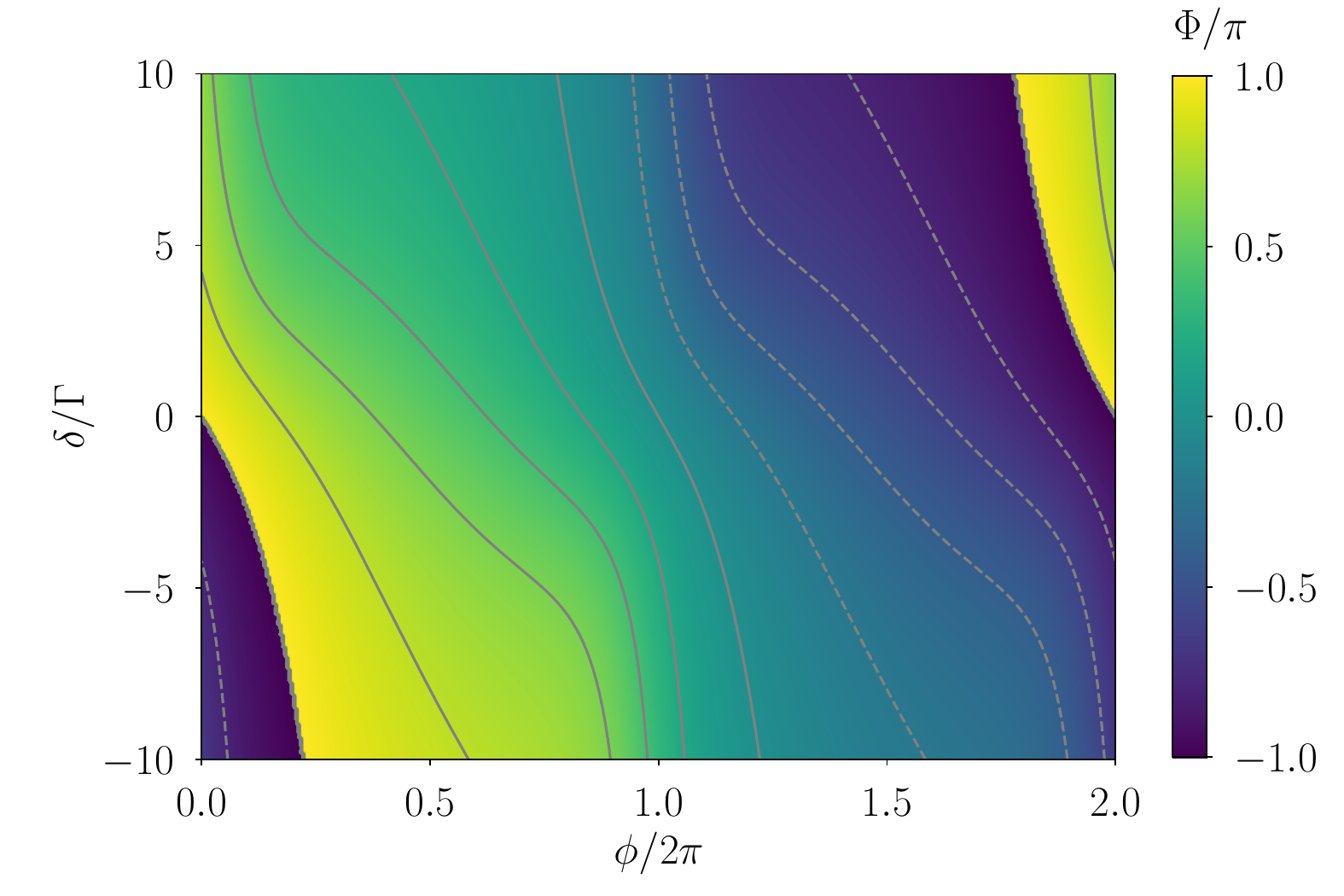}
	\caption{\label{fig:adiabatic}Absolute value $|\mathcal F|$ and phase $\Phi$ of the superconducting pair amplitude on the quantum dot as a function of phase difference $\phi$  and detuning $\delta$. Parameters are $T_\text{L}=0.9T_c$, $T_\text{R}=0.1T_C$, $U=3.6\kBT_c$ and $\Gamma_\text{L}=\Gamma_\text{R}=\Gamma/2$.}
\end{figure}

If the driving frequency is much smaller than the tunneling rates, $\omega\ll\Gamma/\hbar$ the state of the driven quantum dot at a given time $t$ is identical to the stationary state of the undriven system with corresponding system parameters, i.e., the dynamics of the superconducting pair amplitude can be obtained by solving the generalized master equations~\eqref{eq:MEP} and~\eqref{eq:MEI} in the stationary limit $d\vec P/dt=d\vec I/dt=0$ and substituting the time-dependent phase difference $\phi(t)$ or detuning $\delta(t)$. This constitutes a significant simplification compared to the solution of the full, time-dependent master equation which considerably helps in understanding the underlying physics.

The resulting absolute value $|\mathcal F|$ and the phase $\Phi$ of the dot's pair amplitude are shown as a function of the phase difference $\phi$ and the detuning $\delta$ in Fig.~\ref{fig:adiabatic}. We remark that since $\Phi$ is not a physical observable that should be considered relative to the phases of the superconducting reservoirs, it is a $4\pi$-periodic rather than $2\pi$-periodic function of $\phi$.
Let us first consider the situation where the system is driven by a phase difference that increases linearly with time, $\phi(t)=\omega t$. In this case, the absolute value of the pair amplitude $|\mathcal F|$ shows small oscillations with time. The absolute value of the modulation is nearly constant as a function of the detuning $\delta$, however, the relative modulation increases as $\delta$ is tuned away from the particle-hole symmetric point due to the suppression of the proximity effect. The modulation of $|\mathcal F|$ arises because the time-dependent phase difference changes the relative orientation of the pseudospin and the exchange field and, thus, gives rise to a time-dependent modulation of the pseudospin.
The time-dependent phase difference $\phi(t)$ furthermore gives rise to a phase of the pair amplitude on the dot $\Phi$ that decreases with time. This might seem counterintuitive because one might expect that as $\phi_\text{L}=\omega t/2$ increases while $\phi_\text{R}=-\omega t/2$ decreases with time the phase of the pair amplitude on the dot stays nearly constant. However, one has to take into account that the pair amplitude on the dot is a nonequilibrium phenomenon that arises only in the presence of a finite temperature bias which breaks left-right symmetry. This is also the reason why $\Phi$ does not decrease linearly with $\phi$ but rather decreases faster around $\phi=0$ and $\phi=2\pi$.

We now turn to the situation where the system is driven by a time-dependent detuning of the form $\delta=\delta_0+\delta_1\cos\omega t$ where adiabaticity requires a sufficiently small driving amplitude $\delta_1\ll (\kBT)^2/\omega$. It gives rise to oscillations of the absolute value of the dot's pair amplitude $|\mathcal F|$. These oscillations are strongest if the detuning is varied between the particle-hole symmetric point and some finite detuning, i.e., for $\delta_1=\delta_0$ and show little sensitivity to the phase bias $\phi$. The phase of the dot's pair amplitude also shows minor oscillations. These arise because the detuning affects the $z$ component of the exchange field and therefore can accelerate and decelerate the precession of the pseudospin in the $x-y$ plane.

To summarize, we find that for an adiabatic driving of the system the amplitude mode of the pair amplitude can be driven best by a time-dependent detuning $\delta(t)$ while the phase mode is most easily excited by a time-dependent phase bias $\phi(t)$. However, in general both modes are excited at the same time. This constitutes an important difference to the order parameter dynamics in bulk superconductors where only the Higgs mode can be excited at low energies while the Nambu-Goldstone mode is shifted to the plasma frequency.

\subsubsection{\label{sssec:fast}Fast driving}
We now turn to the situation that the driving frequency is much larger than the tunneling rates, $\omega\gg\Gamma/\hbar$. In order to describe this scenario, we expand both the transition rates as well as the density matrix into a Fourier series, 
\begin{align}
	\mathbf W(t)&=\mathbf W_0+\mathbf W_+e^{i\omega t}+\mathbf W_-e^{-i\omega t},\\
	\rho_\text{red}(t)&=\sum_n \rho_n e^{in\omega t},
\end{align}
which allows us to recast the generalized master equation into an infinite hierarchy of coupled equations for the Fourier components of the density matrix
\begin{equation}
	\mathbf W_-\rho_{n+1}+(\mathbf W_0-in\omega)\rho_n+\mathbf W_+\rho_{n-1}=0.
\end{equation}
For driving frequencies $\omega\gg\Gamma/\hbar$, the master equation can be solved approximately by performing a systematic expansion of the density matrix elements in powers of $\Gamma/(\hbar\omega)$ which is equivalent to an expansion in powers of $\mathbf W_\pm$. To lowest order, we obtain the time-averaged density matrix as
\begin{equation}
	\mathbf W_0\rho_0^{(0)}=0.
\end{equation}
The first order correction gives rise to the first harmonics
\begin{equation}\label{eq:rho1}
	\rho_{\pm1}^{(1)}=\frac{1}{\pm i\omega-\mathbf W_0}\mathbf W_\pm \rho_0^{(0)}.
\end{equation}
More generally, we find that the $2n$-th order of the expansion in $\Gamma/(\hbar\omega)$ contributes to all even harmonics up to order $2n$ while the $(2n+1)$-th-order of the expansion in $\Gamma/(\hbar\omega)$ gives contributions to all odd harmonics up to order $2n+1$. As a result, the generation of higher harmonics is suppressed for fast driving. Physically, this is because the dot dynamics becomes too slow to follow the external drive.

\begin{figure}
	\includegraphics[width=\columnwidth]{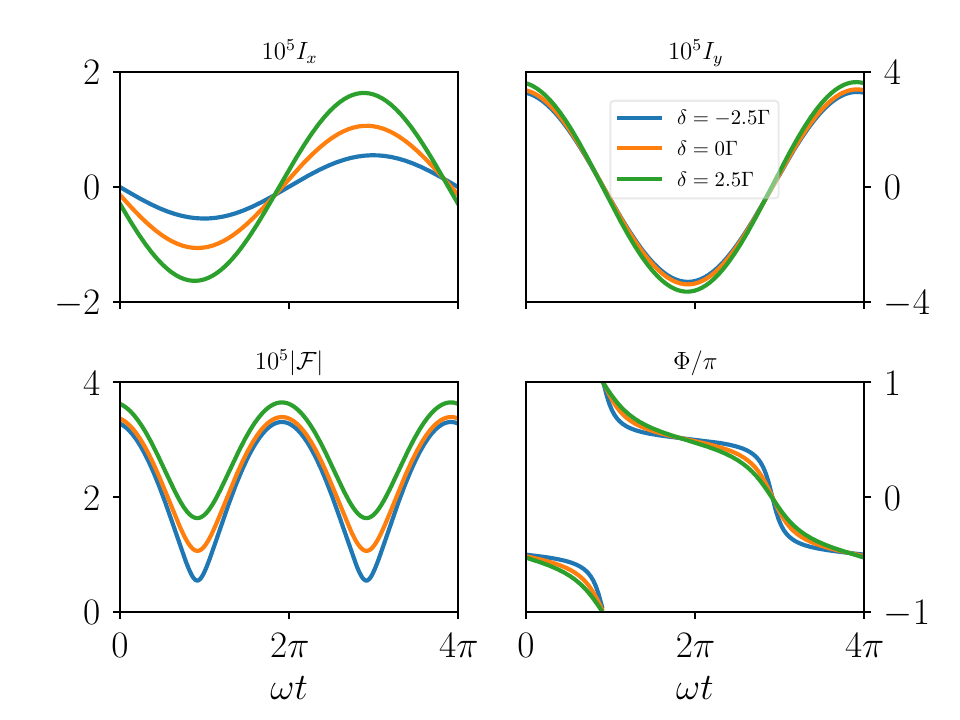}
	\caption{\label{fig:fastdrive}Time dependence of the pseudospin components $I_x$ and $I_y$ as well as of the amplitude $|\mathcal F|$ and the phase $\Phi$ of the pair amplitude on the dot for a driving of the form $\phi(t)=\omega t$. Parameters are $T_\text{L}=0.9T_c$, $T_\text{R}=0.1T_c$, $U=10\kBT_c$, $\Gamma_\text{L}=\Gamma_\text{R}=\Gamma/2$, and $\omega=5\Gamma/\hbar$.}
\end{figure}

Let us now consider the situation where the system is driven by a time-dependent phase difference $\phi(t)=\omega t$. In this case, the matrix $\mathbf W_0$ does not contain any term which accounts for transitions between diagonal and off-diagonal density matrix elements. In consequence, the time-averaged density matrix $\rho_0^{(0)}$ is diagonal such that the average pseudospin components $I_x$ and $I_y$ vanish. A finite pseudospin accumulation in the $x-y$ plane occurs in the first harmonics $\rho_{\pm1}^{(1)}$ to first order in $\Gamma/(\hbar\omega)$. As can be seen in Fig.~\ref{fig:fastdrive}, the resulting amplitude of the pseudospin oscillation is much smaller than in the adiabatic regime. 
Just as in the adiabatic regime, we find that the absolute value of the pair amplitude $\mathcal F$ oscillates with the driving frequency while its phase $\Phi$ and the pseudospin components $I_x$ and $I_y$ oscillate with half the driving frequency. The phase of the pair amplitude $\Phi$ decreases with time which is again linked to the breaking of left-right symmetry by the temperature bias applied between the two superconductors. In contrast to the pseudospin components and the absolute value of the pair amplitude, it does not show a simple sinusoidal time-dependence because it is defined via the ratio of two pseudospin components.

When the system is driven by a time-dependent level detuning $\delta(t)=\delta_0+\delta_1\cos\omega t$, we find a qualitatively similar behavior (not shown). Due to the fast driving, the oscillations of the pseudospin components and the pair amplitude of the quantum dot are small. In contrast to the phase-driven case, we find that the pseudospin and pair amplitude all oscillate with the driving frequency. The different time dependence occurs because for a system driven by a time-dependent detuning, the phase of the dot's pair amplitude can be measured relative to the time-independent phases of the superconducting leads. A second difference to the phase-driven scenario is that the pair amplitude of the quantum dot in general takes a finite time-averaged absolute value when the system is driven by a time-dependent gate voltage.
Similarly to the adiabatic case we find that the amplitude mode of the quantum dot's pair amplitude is excited most easily by a time-dependent level detuning $\delta(t)$ while the phase mode can be excited better with a time-dependent phase difference $\phi(t)$.

\subsubsection{\label{sssec:intermediate}Intermediate driving}
\begin{figure}
    \includegraphics[width=\columnwidth]{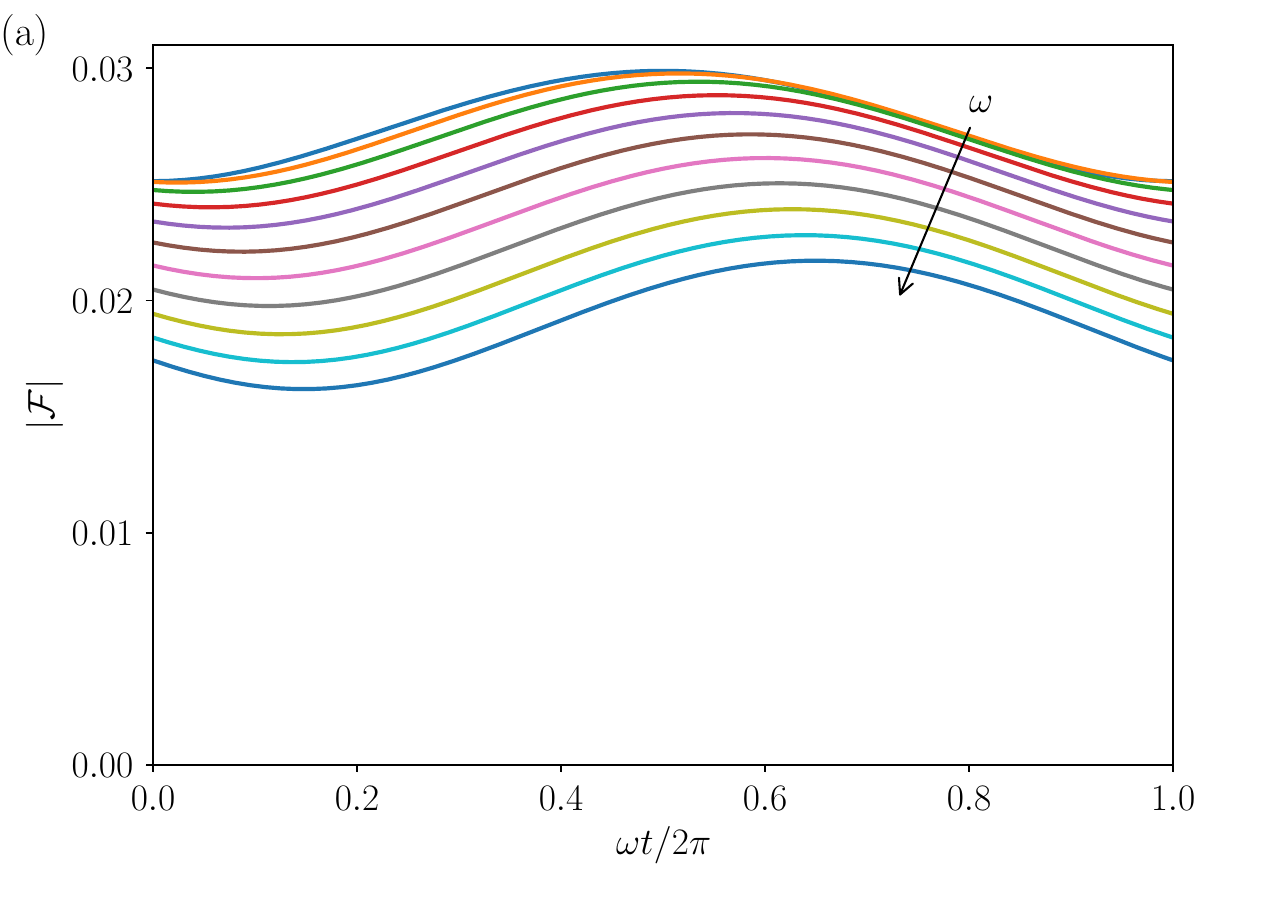}
    \includegraphics[width=\columnwidth]{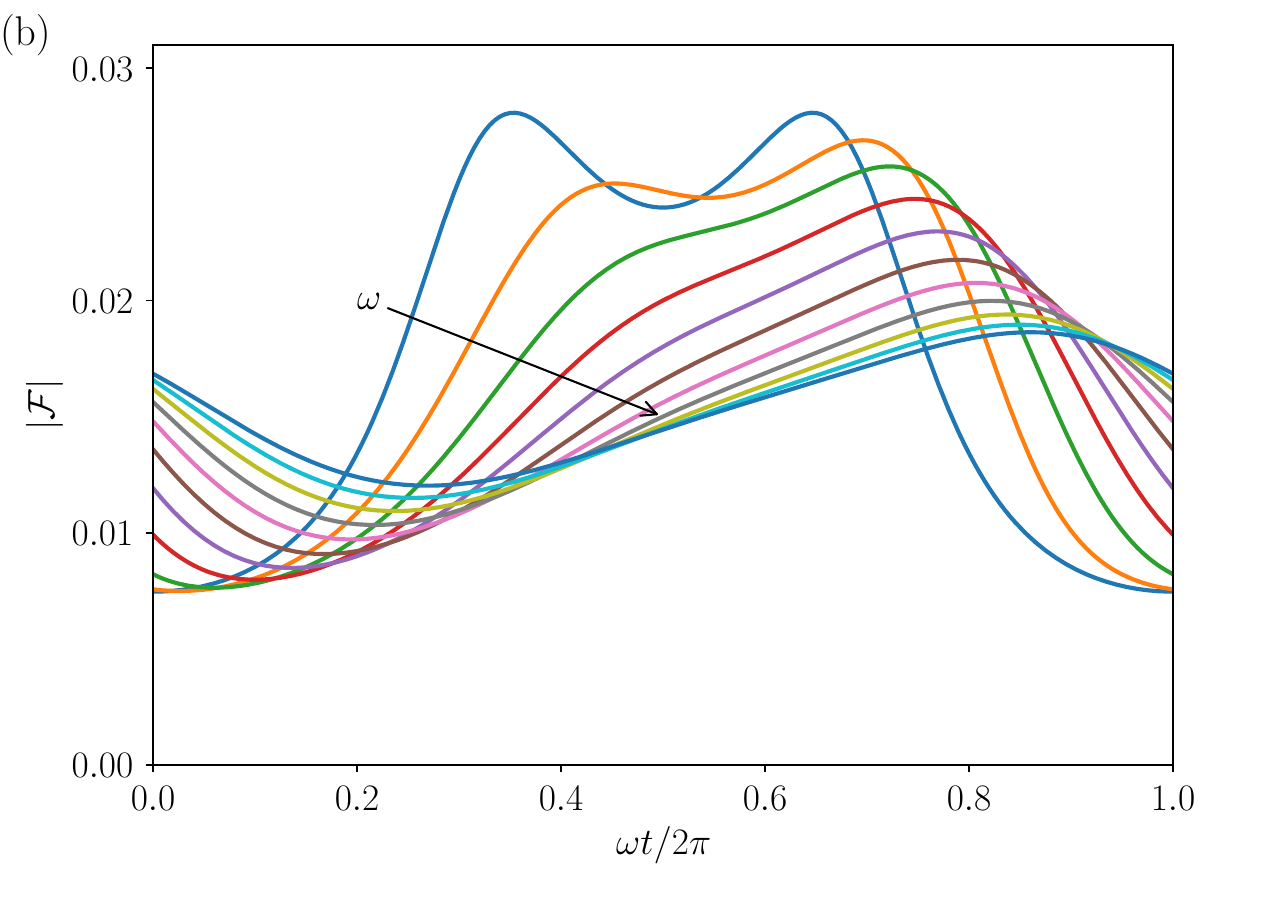}
	\caption{\label{fig:intermediate}Time dependence of the absolute value of the pair amplitude on the quantum dot $|\mathcal F|$ for a system driven by (a) a time-dependent phase difference $\phi(t)=\omega t$ and $\delta=0$ and (b) a time-dependent detuning $\delta(t)=(5/2+10\cos\omega t)\Gamma$ and $\phi=\pi/2$ for driving frequencies from $\omega=0$ to $\omega=2\pi\Gamma/\hbar$ in steps of $0.1\pi\Gamma/\hbar$. Parameters are $T_\text{L}=0.9T_c$, $T_\text{R}=0.3T_C$, $U=4\Delta_0$ and $\Gamma_\text{L}=\Gamma_\text{R}=\Gamma/2$}
\end{figure}

We now turn to the case of intermediate driving $\omega\sim \Gamma/\hbar$ where the time-dependent generalized master equation has to be solved numerically. The resulting time dependence of the absolute value of the superconducting pair amplitude on the quantum dot is shown in Fig.~\ref{fig:intermediate} for a system driven by a time-dependent phase difference $\phi(t)$ and a time-dependent detuning $\delta(t)$, respectively. Compared to the adiabatic case, there is no qualitatively new behavior arising for intermediate driving. Quite generally, we find that the amplitude of the oscillations of $|\mathcal F|$ decreases as $\omega$ is increased in agreement with the results for fast driving. While the generation of higher harmonics can be enhanced for intermediate driving, higher harmonics become suppressed when the driving becomes too fast.

Our numerical solution of the generalized master equation also allows us to address the question up to which driving frequency the adiabatic approximation provides reliable results for the pair amplitude dynamics. Interestingly, we observe that the range of validity depends on which parameter is used to drive the system and on how the parameter changes with time. 

When the system is driven by a time-dependent phase difference $\phi(t)=\omega t$, there is a good agreement between the adiabatic approximation and the full numerical solution of the generalized master equation up to driving frequencies of about $\omega\approx 0.2\Gamma/\hbar$. When the driving frequency is increased further, the pair amplitude becomes suppressed compared to the adiabatic case and is slightly phase-shifted but behaves qualitatively similar to the adiabatic regime otherwise.
When the system is driving by a time-dependent detuning $\delta(t)=\delta_0+\delta_1\cos\omega t$, the deviations between the adiabatic approximation and the numerical solution occur already for lower driving frequencies. They affect in particular the absolute value of the pair amplitude $|\mathcal F|$, cf. Fig.~\ref{fig:intermediate} while they hardly affect its phase $\Phi$ at all (not shown). The deviations from the adiabatic solution are most prominent when the detuning becomes zero or even changes sign during the driving protocol. Physically, this arises because a sign change of $\delta$ implies a sign change of the $z$ component of the exchange field which has a significant impact on the pseudospin dynamics in the $x-y$ plane.

In addition, we find that independent of the precise driving scheme the deviations from adiabaticity are less pronounced for small Coulomb energies $U\approx 2\Delta_0$. In this case, tunneling between the dot and the lead is enhanced by the BCS density of states in the leads such that the effective tunnel coupling can be much larger than $\Gamma_\eta$. As a result, the ratio between the driving frequency and the effective tunnel coupling is reduced and the system is closer to the adiabatic condition.

\section{\label{sec:discussion}Discussion and conclusion}
We have investigated the dynamics of the superconducting pair amplitude of a quantum coupled to two superconducting electrodes under a time-dependent external driving. Using a real-time diagrammatic approach, we have derived a generalized master equation for the reduced density matrix of the quantum dot that accounts for nonequilibrium effects and strong Coulomb interactions exactly and performs a systematic expansion in the tunnel coupling strength. We find that the pair amplitude of the dot can be characterized in terms of a pseudospin which describes coherent superposition of the empty and doubly occupied dot state and which obeys a Bloch-type equation with accumulation and relaxation terms due to electron tunneling and a coherent precession in an effective exchange field due to virtual Andreev tunnel processes.
Since the damping rate is in general one order of magnitude faster than the precession frequency, the relaxation dynamics after a quench is dominated by an exponential decay towards equilibrium where pair amplitude oscillations cannot be observed. This issue can be overcome by a periodic driving of the system which we find  to give rise to self-sustained oscillations of both the absolute value as well as the phase of the pair amplitude. The oscillations are most prominent for adiabatic driving while the amplitude of oscillations is strongly suppressed for fast driving. The oscillations constitute the analogon of the Higgs and Nambu-Goldstone mode in bulk superconductors. Interestingly, driving the system out of equilibrium by a temperature bias is important to reveal the coherent pair amplitude dynamics even in the weak-coupling limit considered here.

We remark that the spin dynamics in a quantum-dot spin valve~\cite{konig_interaction-driven_2003,braun_theory_2004,sothmann_influence_2010}, i.e. a single-level quantum dot coupled to two noncollinearly magnetized ferromagnetic leads can also be analyzed in the framework of Nambu-Goldstone and Higgs mode. In this case, the (real) spin precession caused by an effective exchange field is the analogue of the Nambu-Goldstone mode while the accumulation and relaxation of the spin is connected to the Higgs mode. More generally, quantum dots coupled to reservoirs with spontaneously broken symmetries should always exhibit analogues of Higgs and Nambu-Goldstone modes.

Our results motivate to study the order parameter dynamics in superconductor-quantum dot hybrids in other parameter regimes such as the infinite-gap limit where the superconducting gap provides the largest energy scale in the problem. Furthermore, it is an interesting avenue of future research to link the pair amplitude dynamics of the quantum dot to transport signatures such as charge and heat currents and their respective fluctuations.

\acknowledgments
We acknowledge financial support from the Ministry of Innovation NRW via the ``Programm zur Förderung der Rückkehr des hochqualifizierten Forschungsnachwuchses aus dem Ausland'' and the Deutsche Forschungsgemeinschaft (DFG, German Research Foundation) - Project-ID 278162697 – SFB 1242.

%

\end{document}